\documentclass[11pt]{article}
\pdfoutput=1
\usepackage{fullpage}
\usepackage{xcolor}
\usepackage{cite}
\usepackage{graphicx}
\usepackage{tensor}
\usepackage{amsmath}
\usepackage{amssymb}
\usepackage{subcaption}
\usepackage[utf8x]{inputenc} 
\title{}
\date{}
\renewcommand{\vec}[1]{\mbox{\boldmath$ #1 $}}

\def\beq{\begin{equation}}
\def\eeq{\end{equation}}
\allowdisplaybreaks
\begin{document}
\bibliographystyle{utphys}
\renewcommand{\[}{\begin{equation}\begin{aligned}}
\renewcommand{\]}{\end{aligned}\end{equation}}
\newcommand{\be}{\begin{equation}}
\newcommand{\ee}{\end{equation}}
\newcommand\n[1]{\textcolor{red}{(#1)}} 
\newcommand{\diff}{\mathop{}\!\mathrm{d}}
\newcommand{\lb}{\left}
\newcommand{\rb}{\right}
\newcommand{\f}{\frac}
\newcommand{\pd}{\partial}
\newcommand{\tr}{\text{tr}}
\newcommand{\fdiff}{\mathcal{D}}
\newcommand{\im}{\text{im}}
\let\caron\v
\renewcommand{\v}{\mathbf}
\newcommand{\T}{\tensor}
\newcommand{\R}{\mathbb{R}}
\newcommand{\C}{\mathbb{C}}
\newcommand{\Z}{\mathbb{Z}}
\newcommand{\msbar}{\ensuremath{\overline{\text{MS}}}}
\newcommand{\DIS}{\ensuremath{\text{DIS}}}
\newcommand{\abar}{\ensuremath{\bar{\alpha}_S}}
\newcommand{\bb}{\ensuremath{\bar{\beta}_0}}
\newcommand{\rc}{\ensuremath{r_{\text{cut}}}}
\newcommand{\Nd}{\ensuremath{N_{\text{d.o.f.}}}}
\newcommand{\red}[1]{{\color{red} #1}}
\setlength{\parindent}{0pt}

\titlepage
\begin{flushright}
QMUL-PH-22-24\\
\end{flushright}

\vspace*{0.5cm}

\begin{center}
{\bf \Large Why is the Weyl double copy local in position space?}

\vspace*{1cm} \textsc{Andres Luna$^a$\footnote{andres.luna@nbi.ku.dk}, Nathan Moynihan$^{b,c}$\footnote{nathantmoynihan@gmail.com}
  and Chris D. White$^d$\footnote{christopher.white@qmul.ac.uk} } \\

\vspace*{0.5cm} $^a$Niels Bohr International Academy, Niels Bohr Institute,\\
University of Copenhagen, Blegdamsvej 17, DK-2100, Copenhagen \O, Denmark

\vspace*{0.5cm} $^b$Higgs Centre for Theoretical Physics, School of Physics and Astronomy,\\
The University of Edinburgh, EH9 3FD, Scotland

\vspace*{0.5cm} $^c$School of Mathematics \& Hamilton Mathematics Institute,\\Trinity College Dublin, College Green, Dublin 2, Ireland

\vspace*{0.5cm} $^d$Centre for Theoretical Physics, Department of
Physics and Astronomy, \\
Queen Mary University of London, 327 Mile End
Road, London E1 4NS, UK\\

\end{center}

\vspace*{0.5cm}

\begin{abstract}
The double copy relates momentum-space scattering amplitudes in gauge
and gravity theories. It has also been extended to classical
solutions, where in some cases an exact double copy can be
formulated directly in terms of products of fields in position
space. This is seemingly at odds with the momentum-space origins of the double
copy, and the question of why exact double copies are possible in
position space -- and when this form will break -- has remained
largely unanswered. In this paper, we provide an answer to
this question, using a recently developed twistorial formulation of
the double copy. We show that for certain vacuum type-D solutions, the
momentum-space, twistor-space and position-space double copies amount
to the same thing, and are directly related by integral
transforms. Locality in position space is ultimately a consequence of
the very special form of momentum-space three-point amplitudes, and we
thus confirm suspicions that local position-space double copies are
possible only for highly algebraically-special spacetimes. 
\end{abstract}

\vspace*{0.5cm}

\section{Introduction}
\label{sec:intro}

\begin{figure}
  \begin{center}
    \scalebox{0.8}{\includegraphics{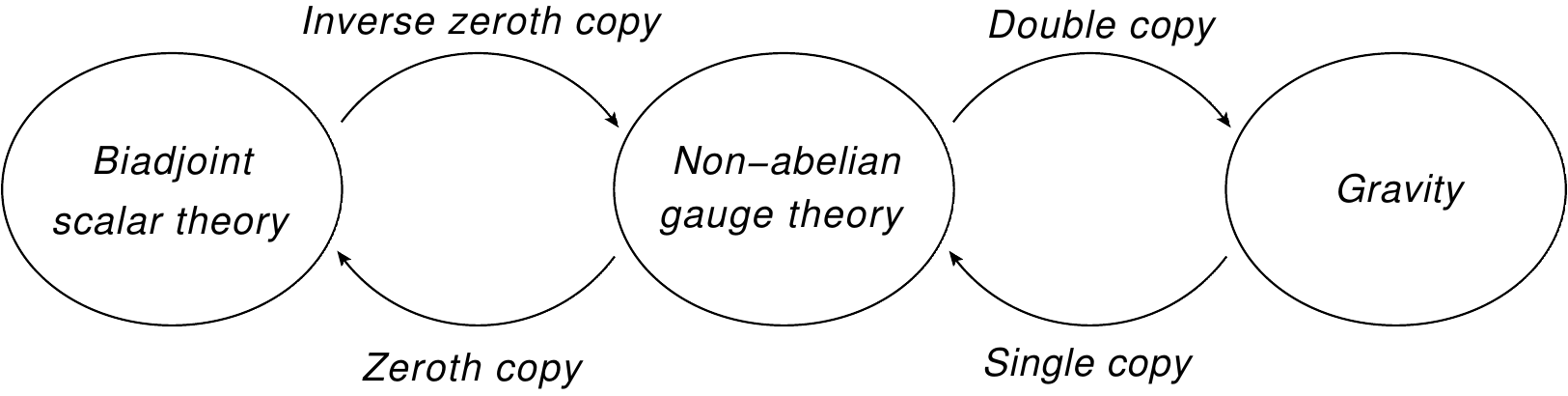}}
    \caption{Different field theories, and the relationships between
      them.}
    \label{fig:theories}
  \end{center}
\end{figure}

Recent years have seen intense study of the relations between
different field theories. One such relation is the {\it double copy},
whose original incarnation relates scattering amplitudes in gauge and
gravity theories~\cite{Bern:2010ue,Bern:2010yg}, and was itself
inspired by earlier work in string theory~\cite{Kawai:1985xq}. Since
then, similar correspondences have been found for amplitudes in a
variety of field theories (see
e.g. ref.~\cite{Bern:2019prr,Borsten:2020bgv,Bern:2022wqg} for recent
reviews). Relevant for the present study is {\it biadjoint scalar
  theory}, consisting of a single scalar field carrying two different
types of colour charge. Its various copy relationships with other
relevant theories are shown in figure~\ref{fig:theories}, and
subsequent work has attempted to establish how generally we are
allowed to interpret this scheme. That it extends beyond scattering
amplitudes was first argued in ref.~\cite{Monteiro:2014cda}, which
showed that certain types of exact classical solution could be copied
between theories (see also refs.~\cite{Didenko:2008va,Didenko:2009td}
for earlier work in a different context), namely those that are of
{\it Kerr-Schild} form in gravity. Whilst algebraically special, this
family of solutions includes cases of astrophysical relevance, such as certain black holes, and cosmologies (e.g. de Sitter space). As
explored in this and many follow-up
works~\cite{Luna:2015paa,Ridgway:2015fdl,Luna:2016due,Bahjat-Abbas:2017htu,Berman:2018hwd,Carrillo-Gonzalez:2017iyj,CarrilloGonzalez:2019gof,Bah:2019sda,Keeler:2020rcv,Easson:2020esh,Alkac:2021seh},
the Kerr-Schild double copy involves products of certain scalar and
vector fields directly in position space. A second exact classical
double copy was formulated in ref.~\cite{Luna:2018dpt}, and further
explored in
refs.~\cite{Sabharwal:2019ngs,Alawadhi:2020jrv,Godazgar:2020zbv,White:2020sfn,Chacon:2021wbr,Chacon:2021hfe,Chacon:2021lox}. It
uses the spinorial formalism of field theory, and is known as the {\it
  Weyl double copy}. Although it looks rather different to the
Kerr-Schild approach, it is equivalent where overlap exists, and also
involves products of spacetime fields directly in position
space. Again, however, the set of solutions that are amenable to being
double-copied is restricted to those that are algebrically special. In
terms of the well-known Petrov classification for gravitational
solutions, the original Weyl double copy was argued to hold for all
vacuum solutions that are of Petrov type D. Particular type-N
solutions have also been explored in ref.~\cite{Godazgar:2020zbv}.\\

It is possible to double copy more-complicated classical gauge theory
solutions to gravity, at the expense of having to work order-by-order
in perturbation theory, using suitable gauge (or other) choices on
both
sides~\cite{Goldberger:2017frp,Goldberger:2017vcg,Goldberger:2017ogt,Goldberger:2019xef,Goldberger:2016iau,Shen:2018ebu,Prabhu:2020avf,Luna:2016hge,Luna:2017dtq,Moynihan:2020ejh,Moynihan:2021rwh,Emond:2021lfy}. Typically,
however, one must formulate such double copies in momentum space,
analogous to how the original double copy for scattering amplitudes
was formulated in the latter. This creates a clear puzzle: even if
direct position-space double copies are restricted to certain classes
of solution only, why should they exist in the first place? The
``natural'' home of the double copy is apparently momentum space, and
one then expects that gravity fields in position space should be
obtainable as {\it convolutions} of spacetime biadjoint and gauge
fields. Indeed, there is an approach that does just
this~\cite{Anastasiou:2014qba,LopesCardoso:2018xes,Anastasiou:2018rdx,Luna:2020adi,Borsten:2020xbt,Borsten:2020zgj,Borsten:2021hua,Godazgar:2022gfw},
which can work in any gauge in principle. Why then, for certain
solutions, can one obtain a {\it product} in position space? This
issue has been addressed recently in ref.~\cite{Monteiro:2021ztt},
which looked in detail at the convolution integrals relating spacetime
gravity solutions to gauge / scalar counterparts, and showed that
these factorise in certain cases into a local product. The
Kerr-Taub-NUT solution was found to be part of this class, linking
with the earlier observations of
refs.~\cite{Monteiro:2014cda,Luna:2015paa}. However, it was noted that
a local product in position space was not possible if one generalises
to solutions that include additional scalar degrees of freedom in the
double copy of pure Yang-Mills theory, such as the dilaton. It will
also be the case that many solutions even in pure gravity do not have
a ``simple'' double copy in position space, and thus
ref.~\cite{Monteiro:2021ztt} is certainly not the last word on this
matter.\\

In this paper, we take a different approach to examining locality of
the Weyl double copy, using various ideas from twistor
theory~\cite{Penrose:1967wn,Penrose:1972ia,Penrose:1968me}. The latter
is a branch of mathematical physics that combines various elements of
algebraic geometry and complex analysis (see
e.g. refs.~\cite{Penrose:1987uia,Penrose:1986ca,Huggett:1986fs,Woodhouse:1985id,Adamo:2017qyl}
for pedagogical reviews), and allows us to visualise certain physical
questions in geometric and / or topological terms. Points in spacetime
are mapped non-locally to objects in an abstract {\it
  twistor space}, and vice versa. This already tells us that issues
relating to locality in spacetime may benefit from viewing them
through a twistorial lens, and indeed a procedure for
``deriving'' the (position-space) Weyl double copy using data in
twistor space has been given in
refs.~\cite{White:2020sfn,Chacon:2021wbr}. At its heart is a
relationship known as the {\it Penrose transform}, that relates
certain contour integrals in twistor space to fields in spacetime. The
integrands of these formulae involve certain twistor ``functions'',
albeit defined only up to equivalence transformations that leave the
integrals invariant. More formally, these quantities are
representatives of cohomology classes, and the twistor double copy
proposed in ref.~\cite{White:2020sfn} is in terms of so-called {\it
  \u{C}ech cohomology}.\\

Representative functions exist for all of the spacetime fields
(scalar, gauge and gravity) entering the Weyl double copy, and
refs.~\cite{White:2020sfn,Chacon:2021wbr} demonstrated that a certain
non-linear product of functions in twistor space corresponds to the
Weyl double copy in position space. This is already intriguing, given
that the map between twistor space and spacetime is
non-local. Furthermore, the non-linear relationship required in
twistor space is obviously at odds with the ability to first perform
equivalence transformations of the various functions that appear. It
seems, then, that particular representatives must be selected for the
twistor double copy to work, but it is not known {\it a priori} what
procedure must be used to systematically fix them. This issue was
explored further in ref.~\cite{Adamo:2021dfg}, which showed that
spacetime data at null infinity could be used to fix representatives
in twistor space, at least for radiative
solutions. Reference~\cite{Chacon:2021lox} considered a different
approach, by first translating from the \u{C}ech cohomology language
to the alternative framework of {\it Dolbeault cohomology}, in which
the Penrose transform integral is interpreted in terms of differential
forms. In Euclidean signature, one may uniquely choose {\it harmonic}
representatives of each required form, upon which the spacetime Weyl
double copy can indeed be shown to correspond to a product structure
in twistor space. However, none of the methods discussed in
refs.~\cite{Adamo:2021dfg,Chacon:2021lox} obviously matches the
original \u{C}ech framework of
refs.~\cite{White:2020sfn,Chacon:2021wbr}, which is arguably simpler
to work with (see e.g. ref.~\cite{Chacon:2021hfe} for a physical
application). Until recently, a simple way of identifying the \u{C}ech
representatives used in refs.~\cite{White:2020sfn,Chacon:2021wbr} has
been lacking. Furthermore, any such procedure should ideally relate to
previously known aspects of the double copy.\\

We can in fact address both the choice of representatives in the
\u{C}ech twistor double copy, and the question of why the Weyl double
copy is local in position space, using the ideas of
refs.~\cite{Guevara:2021yud}. This showed, building on the previous
work of
e.g. refs.~\cite{Monteiro:2020plf,Crawley:2021auj,Monteiro:2021ztt},
how certain classical spacetimes can be obtained from momentum-space
scattering amplitudes. Na\"{i}vely one might think that amplitudes
have nothing to say about classical spacetimes in general: the former
have all external legs on-shell, corresponding to particles that are
radiated to / from past or future null infinity, whereas the latter
are non-radiative in general, and have an off-shell external line
(corresponding to where the spacetime field is being
evaluated). However, as argued in
refs.~\cite{Monteiro:2020plf,Crawley:2021auj,Guevara:2021yud},
non-radiative modes of spacetime fields can indeed probe null infinity
provided one works in (2,2) signature, rather than the usual (1,3)
Lorentzian signature of relativistic quantum field theory. One may
then indeed establish a link between momentum-space scattering
amplitudes in (2,2) signature, and classical solutions in position
space, where one must perform an inverse Fourier transform as
expected.  \\

\begin{figure}
  \begin{center}
    \scalebox{0.8}{\includegraphics{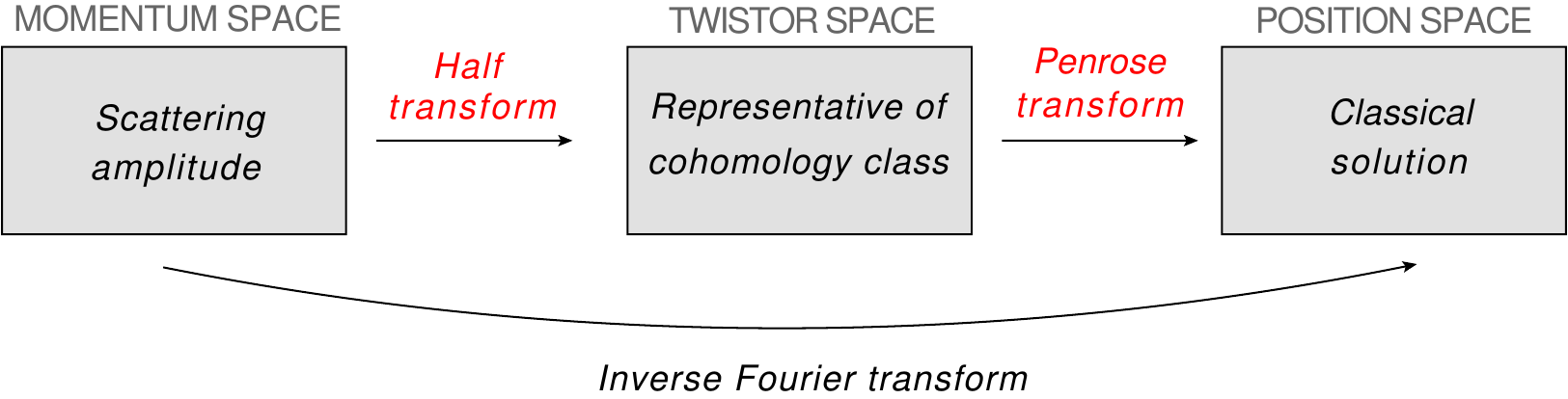}}
    \caption{Scheme proposed in ref.~\cite{Guevara:2021yud} for
      obtaining position-space classical solutions from momentum-space
      scattering amplitudes in (2,2) signature. The form of the ``half
      transform'' is explained in the main text.}
    \label{fig:guevara}
  \end{center}
\end{figure}

Cleverly, ref.~\cite{Guevara:2021yud} takes the equation expressing
classical solutions as inverse Fourier transforms of (2,2) amplitudes,
and splits it into two steps. The first, which we will refer to as the
{\it half-transform}, converts the amplitudes into objects in twistor
space, such that the second step is precisely the Penrose transform
from twistor to position space, which happens to be in the \u{C}ech
language. This scheme is shown in figure~\ref{fig:guevara}, and it is
straightforward to apply it to the classical solutions entering the
twistor double copy considered in
refs.~\cite{White:2020sfn,Chacon:2021wbr}. As mentioned above and
explained in ref.~\cite{Monteiro:2021ztt}, it is known that many of
the type-D solutions entering the Weyl double copy of
ref.~\cite{Luna:2018dpt} can be obtained from scattering amplitudes in
momentum space. The relevant gravity amplitudes can be obtained from
corresponding results in gauge theory, using the double copy as it was
originally formulated. The latter can then be translated into a
relationship between the twistor ``functions'' (representatives of
cohomology classes) living in the middle of figure~\ref{fig:guevara},
which we will show is the twistor double copy of
refs.~\cite{White:2020sfn,Chacon:2021wbr}. Finally, this translates
into the known Weyl double copy in position space, which is equivalent
to the Kerr-Schild double copy where appropriate. We will see the form
of the half transform appearing in figure~\ref{fig:guevara} below, but
it uniquely fixes the representatives in twistor space that are
obtained from given momentum-space amplitudes. Crucially, these
representatives are {\it precisely} those \u{C}ech representatives
that appear in the original twistor double copy. Thus, the latter is a
true consequence of the double copy for scattering amplitudes, and
this even suggests how the twistor approach may be extended (e.g. by
translating higher-order amplitudes into the twistor language). \\

In summary, by fleshing out the details of figure~\ref{fig:guevara},
we firmly establish the complete equivalence of the BCJ double copy
for three-point amplitudes~\cite{Bern:2010ue,Bern:2010yg}, the twistor
double copy of refs.~\cite{White:2020sfn,Chacon:2021wbr}, and the type
D Weyl double copy of ref.~\cite{Luna:2018dpt}, at least for those
type-D solutions where corresponding amplitudes are known. This is
itself puzzling: the maps between all three spaces are non-local, and
yet the double copy takes a manifestly local form in all three! We
will be able to ascertain why this is the case, and it will only turn
out to be true due to the highly-special form of the relevant
three-point amplitudes in momentum space. Not only does this settle
the question of why local position-space double copies are possible,
but it also confirms that such situations are not generic, but rely on
very special circumstances. \\

The structure of our paper is as follows. In section~\ref{sec:review},
we review ideas relating to the twistor double copy of
refs.~\cite{White:2020sfn,Chacon:2021wbr}. In
section~\ref{sec:guevara}, we apply the methods of
ref.~\cite{Guevara:2021yud} to demonstrate that scattering amplitudes
in momentum space can be used to pick out the \u{C}ech cohomology
representatives entering the twistor double copy. In
section~\ref{sec:simple}, we explain why locality of the double copy
is simultaneously manifest in momentum, twistor and position space,
for type-D solutions. Finally, we discuss our results and conclude in
section~\ref{sec:discuss}.

\section{The twistor double copy}
\label{sec:review}

In this section, we review various aspects of the twistor double copy
introduced in refs.~\cite{White:2020sfn,Chacon:2021wbr}, both to make
the paper reasonably self-contained, and also to set up notation
needed for what follows. As mentioned above, the twistor double copy
reproduces the Weyl double copy in position space, so we must first
recap the definition of the latter.

\subsection{Spinors and the Weyl double copy}
\label{sec:spinors}
The Weyl double copy relies on the spinorial formalism of field
theory, in which all equations of motion are written in terms of
two-component Weyl spinors $\pi_A$, or conjugate spinors $\pi_{A'}$,
and their multi-index generalisations. Here the indices $A$,
$A'\in\{0,1\}$, where indices may be raised and lowered using the
two-dimensional Levi-Civita symbols:
\begin{equation}
  \pi_A=\epsilon_{AB}\pi^B,\quad \pi^B=\pi_A \epsilon^{AB},
  \label{piAlower}
\end{equation}
where
\begin{equation}
  \epsilon_{AB}\epsilon^{CB}=\delta^C_A,\quad \epsilon_{01}=1.
  \label{epsilondef}
\end{equation}
Similar equations hold for raising and lowering indices of conjugate
spinors, but using $\epsilon_{A'B'}$ etc. such that
\begin{equation}
  \epsilon_{A'B'}\epsilon^{C'B'}=\delta^{C'}_{A'},\quad \epsilon_{0'1'}=1.
  \label{epsilondef2}
\end{equation}
To convert between spacetime indices\footnote{Throughout the paper, we
use lower-case Latin letters, upper-case Latin letters and Greek
letters for tensor, spinor and twistor indices respectively.} and
spinor indices, one may use the {\it Infeld-van-der-Waerden symbols}
$\{\sigma^a_{AA'},\sigma_a^{AA'}\}$. Given that we wish to make
contact with refs.~\cite{Monteiro:2020plf,Guevara:2021yud}, we will
work in a (2,2) spacetime signature throughout, for which a suitable
choice for the Infeld-van-der-Warden symbols
is~\cite{Monteiro:2020plf}:
\begin{equation}
  \sigma^a=(1,i\sigma_y,\sigma_z,\sigma_x),
  \label{sigmamudef}
\end{equation}
where $1$ denotes the $2\times 2$ identity matrix, and we have used
the Pauli matrices
\begin{equation}
  \sigma_x=\left(\begin{array}{cc} 0& 1 \\ 1 & 0\end{array}\right),\quad
  \sigma_y=\left(\begin{array}{cc} 0& -i \\ i & 0\end{array}\right),\quad
  \sigma_z=\left(\begin{array}{cc} 1& 0 \\ 0 & -1\end{array}\right).    
  \label{Pauli}
\end{equation}
According to these conventions, a spacetime 4-vector has the following
spinorial translation:
\begin{equation}
  V_{AA'}\equiv V_a \,\sigma^a_{AA'}=
  \left(\begin{array}{cc}V_0+V_2 & V_1+V_3\\
    V_3-V_1 & V_0-V_2\end{array}\right),
    \label{Vspinor}
\end{equation}
from which one obtains the handy formula
\begin{equation}
  V\cdot W=\frac{1}{2}V_{AA'} W^{AA'}.
  \label{VdotW}
\end{equation}
The determinant of the matrix in eq.~(\ref{Vspinor}) is
\begin{equation}
  |V_{AA'}|=\left(V_0^2+V_1^2-V_2^2-V_3^2\right)=V^2,
  \label{detV}
\end{equation}
which therefore vanishes for null vectors ($V^2=0$). This in turn
implies that one may factorise the matrix into an outer product of two
spinors:
\begin{equation}
  V_{AA'}=\pi_A \tilde{\pi}_{A'},\quad V^2=0.
\label{Vfac}
\end{equation}
A consequence of the limited range of spinorial indices is that all
multi-index spinors can be decomposed into products of fully-symmetrised spinors, and Levi-Civita symbols. As an example, the
spinorial translation of the field strength tensor in electromagnetism
can be written as follows:
\begin{equation}
  F_{ab}\rightarrow F_{AA'BB'}=\phi_{AB}\epsilon_{A'B'}+
  \bar{\phi}_{A'B'}\epsilon_{AB}.
  \label{Fmunu}
\end{equation}
Here the symmetric spinors $\phi_{AB}$ and $\bar{\phi}_{A'B'}$
respectively represent the anti-self-dual and self-dual degrees of
freedom in the electromagnetic field. Another important case is that
of vacuum gravitational solutions, for which the Riemann curvature
tensor $R_{abcd}$ reduces to the {\it Weyl tensor},
with spinorial translation
\begin{equation}
  C_{abcd}\rightarrow \phi_{ABCD}\epsilon_{A'B'}
  \epsilon_{C'D'}+\bar{\phi}_{A'B'C'D'}\epsilon_{AB}\epsilon_{CD}.
  \label{Weylspinor}
\end{equation}
Again, (un-)barred quantities correspond to the (anti-)self-dual parts
of the field. The various quantities appearing in eqs.~(\ref{Fmunu},
\ref{Weylspinor}) obey special cases of the general {\it massless free
  field equations}
\begin{equation}
  \nabla^{AA'}\phi_{AB\ldots C}=0,\quad \nabla^{AA'}\bar{\phi}_{A'B'
    \ldots C'}=0,
  \label{masslessfreefield}
\end{equation}
with $\nabla^{AA'}$ the spinorial translation of the covariant
derivative. A spin-$n$ spacetime field leads to a multi-spinor field
with $2n$ indices. Following convention, we will refer to the $n=1$
and $n=2$ cases as electromagnetic and Weyl spinors respectively.\\

Again due to the two-valued nature of spinor indices, it turns out
that all symmetric multi-index spinors can be factorised into a
symmetrised product of 1-index {\it principal spinors}. For
electromagnetic and Weyl spinors, this takes the explicit form
\begin{equation}
  \phi_{AB}=\alpha_{(A}\beta_{B)},\quad
  \phi_{ABCD}=\alpha_{(A}\beta_B\gamma_C\delta_{D)}.
  \label{principal}
\end{equation}
We may then classify solutions of electromagnetism and gravity into
qualitatively different types, according to the degeneracy of their
principal spinors. Electromagnetic fields are referred to as
(non-)null, if their principal spinors are (not) proportional. There
are many more possibilities for gravity solutions, which we list in
table~\ref{tab:Petrov}. This is known as the {\it Petrov
  classification}, and different patterns of principal spinors
constitute different {\it Petrov types}. Given a principal spinor
$\xi_A$, we may take its complex conjugate $\tilde{\xi}^{A'}$ and form
a spacetime vector according to
\begin{equation}
  x^a=\sigma^a_{AA'}\xi^A\tilde{\xi}^{A'},
  \label{ximu}
\end{equation}
which will be null in accordance with eq.~(\ref{Vfac}). Thus,
principal spinors translate to so-called {\it principal null
  directions} in the tensorial language. \\
\begin{table}
\begin{center}
\begin{tabular}{c|c}
Weyl type & Petrov label\\
\hline
$\{1,1,1,1\}$ & I\\
$\{2,1,1\}$ & II \\
$\{3,1\}$ & III\\
$\{4\}$ & N\\
$\{2,2\}$ & D\\
$\{-\}$ & O 
\end{tabular}
\caption{Different types of Weyl spinor classified by: (i) the
  pattern of degenerate principal null directions; (ii) the
  equivalent Petrov type.}
\label{tab:Petrov}
\end{center}
\end{table}

Given certain electromagnetic spinors
$\{\phi^{(1)}_{AB},\phi^{(2)}_{AB}\}$ and a scalar field $\phi$, the
{\it Weyl double copy} states that
\begin{equation}
  \phi_{ABCD}=\frac{\phi^{(1)}_{(AB}\phi^{(2)}_{CD)}}{\phi}
  \label{WeylDC}
\end{equation}
is a Weyl spinor, corresponding to a particular gravitational
solution~\cite{Luna:2018dpt}. The original incarnation of this formula
applied to only those cases in which
$\phi^{(1)}_{AB}=\phi^{(2)}_{AB}$, and was argued to hold for arbitrary
Petrov type-D solutions. Further work has established the existence of
{\it mixed} Weyl double copies with
$\phi^{(1)}_{AB}\neq\phi^{(2)}_{AB}$, with applications to certain
type N solutions~\cite{Godazgar:2020zbv}, as well as other Petrov
types at linearised level
only~\cite{White:2020sfn,Chacon:2021wbr}. Other implications have been
explored in
refs.~\cite{Alawadhi:2020jrv,Easson:2021asd,Han:2022ubu,Han:2022mze,Godazgar:2021iae},
and a novel three-dimensional counterpart of the Weyl double copy (the
{\it Cotton double copy}) has recently been proposed in
refs.~\cite{Emond:2022uaf,Gonzalez:2022otg}. Note that
eq.~(\ref{WeylDC}) involves products of fields in position space,
which is mysterious given that the original double copy for scattering
amplitudes is naturally expressed in momentum space. This implies that
one should expect {\it convolutions} of fields in position space, and
indeed
refs.~\cite{Anastasiou:2014qba,LopesCardoso:2018xes,Anastasiou:2018rdx,Luna:2020adi,Borsten:2020xbt,Borsten:2020zgj,Borsten:2021hua}
imply that this will be true in general for classical fields. For
specific solutions, ref.~\cite{Monteiro:2021ztt} has pointed out that
the mathematical properties of the relevant convolution integrals are
such that products of fields can indeed be made manifest in both
position and momentum space. Here, we shed more light on this issue by
using the twistor methods outlined below.

\subsection{The twistor double copy}

We may define a twistor to be a composite object containing two
spinors of opposite chirality
\begin{equation}
  Z^\alpha=(\lambda_{A},\mu^{A'}),
  \label{twistor}
\end{equation}
whose components satisfy the {\it incidence relation}\footnote{One way
to interpret eqs.~(\ref{twistor}, \ref{incidence}) is that the spinors
in $Z^\alpha$ characterise independent solutions of the {\it twistor
  equation} $\nabla_A^{(A'}\Lambda^{B')}=0$, for some spinor field
$\Lambda^{B'}$. The incidence relation then arises by defining the
location of a twistor in spacetime by $\Lambda^{A'}=0$. See
e.g. ref.~\cite{Penrose:1986ca}.}
\begin{equation}
  \mu^{A'} = x^{AA'}\lambda_A.
  \label{incidence}
\end{equation}
Twistor space $\mathbb{T}$ consists of all objects of the form of
eq.~(\ref{twistor}). However, eq.~(\ref{incidence}) is invariant under
rescalings of both sides (and thus the twistor of eq.~(\ref{twistor}))
by a common factor $\lambda$. Thus, twistors satisfying the incidence
relation are points in {\it projective twistor space}
$\mathbb{PT}$. Unless otherwise stated, we will consider complexified
flat spacetime in (2,2) signature, with Cartesian line element
\begin{equation}
  ds^2=dt^2+dx^2-dy^2-dz^2,\quad t,x,y,z\in\mathbb{C}.
  \label{ds2}
\end{equation}
All twistor components are then real, and it is straightforward
to ascertain that the incidence relation comprises a {\it non-local}
map between $\mathbb{PT}$ and spacetime. For example, a given point in
$\mathbb{PT}$ is associated with all spacetime points satisfying
eq.~(\ref{incidence}), which are of the form
\begin{equation}
  x^{AA'}=x_0^{AA'}+\lambda^A\alpha^{A'}.
  \label{xAA'sol}
\end{equation}
Here $x_0^{AA'}$ is a fixed point in spacetime, and $\lambda_{A}$ is
also fixed for a given point in $\mathbb{PT}$. Equation~(\ref{Vfac})
then reveals that the second term on the right-hand side generates a
null direction in spacetime for a given $\alpha^{A'}$. Thus, varying
$\alpha^{A'}$ generates a set of null directions, and thus a {\it null
  plane} in (complex) spacetime: a plane such that all tangent vectors
are null. These are called $\alpha$-planes, and we may also note that
were we to restrict to a real spacetime in Lorentzian signature, we
would obtain a null geodesic (line) rather than a null plane in
spacetime. To see this, note that in Lorentzian signature,
eq.~(\ref{Vspinor}) would be replaced with 
\begin{equation}
  V_{AA'}\equiv V_a \,\sigma^a_{AA'}=
  \left(\begin{array}{cc}V_0+V_2 & V_3+iV_1\\
    V_3-iV_1 & V_0-V_2\end{array}\right),
    \label{Vspinor2}
\end{equation}
where all coordinates $V_a$ are real. This in turn implies that the
spinor $\alpha^{A'}$ appearing in eq.~(\ref{xAA'sol}) must be related
to the complex conjugate $\tilde{\lambda}^{A'}$ of $\lambda_{A}$ up to
a constant factor:
\begin{displaymath}
  \alpha^{A'}\propto \tilde{\lambda}^{A'}.
\end{displaymath}
This picks out the unique null direction specified by $\lambda_{A}$
(which is fixed for a given twistor), as required.\\

So much for the map from $\mathbb{PT}$ to complex spacetime. To go the
other way round, we may note that a point in twistor space has (from
eq.~(\ref{twistor})) 4 complex degrees of freedom, reducing to 3 if we
consider $\mathbb{PT}$. The incidence relation of
eq.~(\ref{incidence}) then provides a further 2 constraints, so that a
fixed point in spacetime constitutes a single degree of freedom, or
(complex) line, in $\mathbb{PT}$. We can take points on this line to
be specified by the twistor components $\pi_{A'}$ which, given the
projective nature of the space, we may parametrise according to either
\begin{equation}
  \pi_{A'}=(1,\xi)\quad\rm{or} \quad \pi_{A'}=(\eta,1),\quad
  \xi,\eta\in\mathbb{C}.
  \label{piparam}
\end{equation}
These define two coordinate patches covering a Riemann sphere, which
has a nice geometric interpretation in the real Lorentzian case. Given
$\pi_{A'}$ corresponds to a null direction emanating from the fixed
point $x_0^{AA'}$, the Riemann sphere corresponding to a fixed
spacetime point constitutes the {\it celestial sphere} of all possible
null directions from $x_0^{AA'}$ (up to reparametrisations). We will
refer to the Riemann sphere corresponding to a specific 
spacetime point $x^{AA'}$ as $X$ in what follows.\\

\begin{figure}
  \begin{center}
    \scalebox{0.8}{\includegraphics{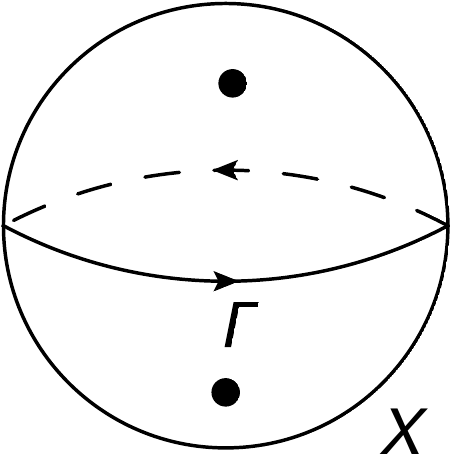}}
    \caption{The Penrose transform involves the integral of a twistor
      ``function'' $f(Z^\alpha)$ around a contour $\Gamma$ on the
      Riemann sphere $X$ corresponding to a given spacetime point
      $x$. For a non-zero result, there must be at least one pole on
      either side of the contour.}
    \label{fig:Penrosepoles}
  \end{center}
\end{figure}

Given the twistor of eq.~(\ref{twistor}), one may also define a {\it
  dual twistor}
\begin{equation}
  W_\alpha=(\tilde{\mu}^A,\tilde{\lambda}_{A'}).
  \label{dualtwistor}
\end{equation}
This allows one to define an inner product between (dual) twistors:
\begin{equation}
  Z^\alpha W_\alpha=\tilde{\mu}^A\lambda_A+\mu^{A'}\tilde{\lambda}_{A'}.
  \label{innerprod}
\end{equation}
As discussed in the introduction, a key result of twistor theory is
the fact that solutions of the massless free field equations of
eq.~(\ref{masslessfreefield}) can be represented using certain
integral formulae in $\mathbb{PT}$. More specifically, the {\it
  Penrose transform} expresses the self-dual part of a spin-$n$
field as
\begin{equation}
  \phi_{AB\ldots C}(x)=\frac{1}{2\pi i}\oint_\Gamma
  \lambda_{E}d\lambda^{E}\lambda_{A}\lambda_{B}\ldots
  \lambda_{C}[\rho_x f(Z^\alpha)].
  \label{Penrose}
\end{equation}
The right-hand contains a holomorphic ``function'' of a single twistor
variable $f(Z^\alpha)$, where the symbol $\rho_x$ denotes restriction
to the Riemann sphere $X$ corresponding to spacetime point
$x^{AA'}$. The remaining integrand contains factors of the spinor
$\lambda_{A}$ that enters the twistor $Z^\alpha$ (as in
eq.~(\ref{twistor})), and the contour $\Gamma$ is such as to separate
any poles of $f(Z^\alpha)$ on $X$. An example is shown in
figure~\ref{fig:Penrosepoles}, and for the contour integral to give a
non-zero answer, one must clearly have at least one pole on either
side of the contour $\Gamma$.
For a given spacetime field, the function $f(Z^\alpha)$ is not
uniquely defined: it may be subjected to equivalence transformations
of the form
\begin{equation}
  f(Z^\alpha)\rightarrow f(Z^\alpha)+f_N(Z^\alpha)+f_S(Z^\alpha),
  \label{ftrans}
\end{equation}
where $f_N(Z^\alpha)$ ($f_S(Z^\alpha)$) has poles only in the northern
(southern) hemisphere of $X$, without changing the result of the
contour integral. Mathematically speaking, $f(Z^\alpha)$ is a
representative of a {\it cohomology class}, and one may formalise this
discussion in terms of {\it \u{C}ech cohomology groups}, which are
themselves approximations to {\it sheaf cohomology groups}, as
discussed in
refs.~\cite{Eastwood:1981jy,Penrose:1986ca,Huggett:1986fs}. An
alternative formulation exists using the language of differential
forms and {\it Dolbeault cohomology}, as reviewed e.g. in
ref.~\cite{Adamo:2017qyl}, and a recent discussion of a comparison
between the two approaches can be found in
ref.~\cite{Chacon:2021lox}. We will use the \u{C}ech approach
throughout, and the next point we need to note is that the requirement
that eq.~(\ref{Penrose}) make sense as an integral in projective
twistor space imposes a restriction on $f(Z^\alpha)$. That is, the
integral must give the same answer under rescalings
$Z^\alpha\rightarrow \lambda Z^\alpha$, which can only be true if one
has
\begin{equation}
  f(Z^\alpha)\rightarrow \lambda^{-2n-2}f(Z^\alpha),
  \label{frescale}
\end{equation}
if there are $2n$ indices on the left-hand side of the Penrose
transform. That is, a spin-$n$ field corresponds to a cohomology
representative $f(Z^\alpha)$ with homogeneity $(-2n-2)$. For scalar,
electromagnetic and gravitational fields respectively, this implies
homogeneities $-2$, $-4$ and $-6$. We have here addressed the case of
the self-dual part of a massless free field. The anti-self-dual part
can be obtained using an alternative Penrose transform in terms of
dual twistors, as discussed in
e.g. refs.~\cite{Penrose:1986ca,Huggett:1986fs}.\\

We are now able to state the twistor expression of the Weyl double
copy that first appeared in
refs.~\cite{White:2020sfn,Chacon:2021wbr}. Given certain cohomology
representatives $f_{-2}(Z^\alpha)$, $f_{-4}^{(1)}(Z^\alpha)$ and
$f_{-4}^{(2)}(Z^\alpha)$, where the subscript denotes the homogeneity,
one may construct a homogeneity $-6$ representative via the product
\begin{equation}
  f_{-6}(Z^\alpha)=\frac{f_{-4}^{(1)}(Z^\alpha)f_{-4}^{(2)}(Z^\alpha)}
  {f_{-2}(Z^\alpha)}.
  \label{fgrav}
\end{equation}
By the above remarks, this will correspond to a gravitational
field. However, the constituent ``functions'' on the right-hand side
correspond to a pair of electromagnetic fields, and a scalar. There
must then be a relationship between the corresponding spacetime
fields, and refs.~\cite{White:2020sfn,Chacon:2021wbr} showed that one
may choose representatives such that this spacetime relationship is
precisely the type-D Weyl double copy. To do this, one may rely on the
observation made in ref.~\cite{Penrose:1986ca}, that -- for
representatives $f(Z^\alpha)$ that involve only two poles -- a pole of
order $m$ in twistor space leads to a $(n-m+1)$-fold degenerate
principal spinor in spacetime, for $n$ the spin. A type-D solution has
two 2-fold degenerate principal spinors, so that it may be generated
using a cohomology representative of form
\begin{equation}
  f_{-6}(Z^\alpha)=\left[Q^{\alpha\beta} Z^\alpha Z^\beta\right]^{-3},
  \label{f-6Q}
\end{equation}
where $Q^{\alpha\beta}$ is a constant twistor. Likewise, one may
generate scalar and electromagnetic fields via the choices
\begin{equation}
  f_{-2}(Z^\alpha)=\left[Q^{\alpha\beta} Z^\alpha Z^\beta\right]^{-1},\quad
  f_{-4}^{(1)}(Z^\alpha)=f_{-4}^{(2)}(Z^\alpha)
  =\left[Q^{\alpha\beta} Z^\alpha Z^\beta\right]^{-2}.
  \label{f-4Q}
\end{equation}
It is easily checked that these representatives obey
eq.~(\ref{fgrav}). Furthermore, choosing different forms for
$Q^{\alpha\beta}$ is sufficient to map out the complete space of
vacuum type-D solutions~\cite{Haslehurst}. \\

As remarked above, the quantities $f(Z^\alpha)$ entering the Penrose
transform integral are representatives of cohomology classes which, in
more pedestrian terms, amount to functions defined only up to the
equivalence transformations of eq.~(\ref{ftrans}). The product of
eq.~(\ref{fgrav}), needed to reproduce the Weyl double copy in
position space, is clearly incompatible with the ability to first
perform equivalence transformations of the gauge and / or scalar
functions. Furthermore, this is unavoidable, given that the
combination of twistor ``functions'' required by the double copy is
necessarily non-linear. It seems, then, that the product-like nature
of the twistor space double copy is only possible if special
representatives of each cohomology class are chosen, and it is not
clear a priori what these representatives should be.\\

Reference~\cite{Adamo:2021dfg} was the first work to provide a
potential solution to this issue, at least for radiative spacetimes
that can be fully defined by specifying data at future null
infinity. A certain procedure exists~\cite{MasonTN} for using this
data to fix twistor representatives of spacetime fields, in the
Dolbeault cohomology framework alluded to
above. Reference~\cite{Adamo:2021dfg} then argued that a twistorial
double copy naturally emerges for these
representatives. Reference~\cite{Chacon:2021lox} considered both the
\u{C}ech and Dolbeault languages, first showing that one may translate
the original \u{C}ech double copy of
refs.~\cite{White:2020sfn,Chacon:2021wbr} into the Dolbeault approach,
albeit subject to the same conceptual puzzle regarding how to pick
cohomology representatives. It then showed that, for solutions in
Euclidean signature, established techniques imply that there are
unique choices of Dolbeault representative -- namely those that are
{\it harmonic} differential forms~\cite{Woodhouse:1985id} -- such that
the Weyl double copy in position space yields a product structure in
twistor space. Whilst this is encouraging, it is not known how to
directly relate these representatives to those in the \u{C}ech
language, nor is it known what the harmonic condition implies for the
latter. It also not known whether this procedure can be directly
related to that of ref.~\cite{Adamo:2021dfg}. From a mathematical
point of view, it is not clear whether the different double copy
procedures in twistor space amount to the same double copy in position
space, or a set of physically distinct double copy procedures. If the
latter turns out the case, one can then ask which twistor double copy,
if any, corresponds to the original double copy for scattering
amplitudes. We provide an answer to this question in the following
section.

\section{Cohomology representatives from scattering amplitudes}
\label{sec:guevara}

Above, we have posed the question of which twistor double copy
procedure, if any, can be related to the double copy for scattering
amplitudes. In fact, the recent developments of
refs.~\cite{Monteiro:2020plf,Guevara:2021yud}, allow us to precisely
answer this question. We begin by showing how the scheme of
figure~\ref{fig:guevara} can be made precise.\\

\subsection{From scattering amplitudes to twistor space}
\label{sec:halftrans}
\begin{figure}
  \begin{center}
    \scalebox{0.6}{\includegraphics{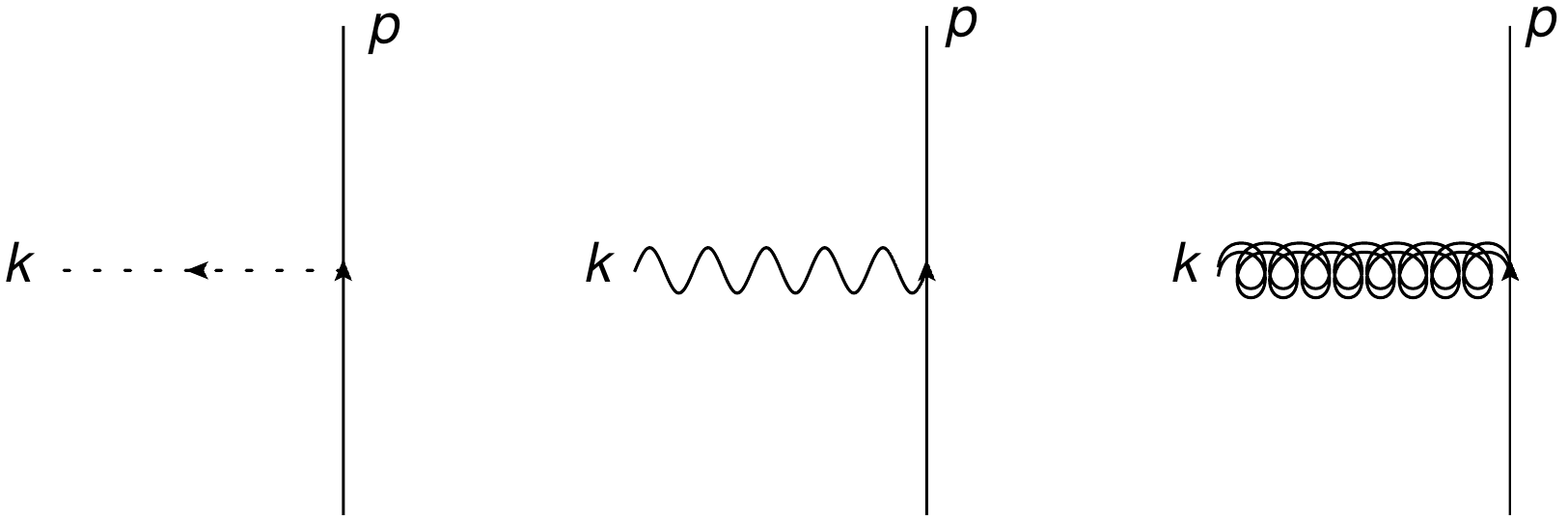}}
    \caption{Three-point amplitudes for the emission of a scalar,
      photon or graviton from static sources.}
    \label{fig:sources}
  \end{center}
\end{figure}

Let us consider three-point amplitudes for emission of a scalar,
photon or graviton from a static source particle, as shown in
figure~\ref{fig:sources}.
Following ref.~\cite{Guevara:2021yud}, we may write the spinorial
translation of the radiation momentum $k^\mu$ as
\begin{equation}
  k_{AA'}=\omega \lambda_A\tilde{\lambda}_{A'}+\xi q_{AA'},
  \label{kAA'}
\end{equation}
where $\omega=k^0$ is the energy, and $\lambda_A$,
$\tilde{\lambda}_{A'}$ are dimensionless spinors. In what follows, we
will parametrise these by 
\begin{equation}
  \lambda_A=(1,z),\quad \tilde{\lambda}_{A'}=(1,\tilde{z})
  \label{lambdaparam}
\end{equation}
for $z,\tilde{z}\in\mathbb{C}$, such that we may think of each spinor
as defining a point on a Riemann sphere, whose meaning will be
clarified shortly\footnote{To cover the Riemann sphere, we would need
to consider a second coordinate patch in which $\lambda_1\neq 0$,
$\tilde{\lambda}_{1'}\neq 0$. We will not need to consider this explicitly
for our purposes.}. We have also introduced a null reference vector
$q_{AA'}$ in eq.~(\ref{kAA'}), such that $\xi$ parametrises the
off-shellness of $k^\mu$.\\

Denoting the amplitude for spin-$n$ radiation by
${\cal A}^{(n)}_\pm$, where $\pm$ denotes the helicity of the emitted
boson as appropriate, one may obtain the classical unprimed spinor
field for the emitted radiation via the integral formula:
\begin{align}
  \phi_{A_1A_2\ldots A_{2n}}&=N_n\,{\rm Re}\int d\Phi(k)\,2\pi\delta (2p\cdot k)
  \omega^n\lambda_{A_1}\lambda_{A_2}\ldots \lambda_{A_{2n}}
         e^{-ik\cdot x}{\cal A}^{(n)}_{+},
         \label{phiFourier}
\end{align}
where \begin{equation}
  d\Phi(k)\equiv \frac{d^4 k}{(2\pi)^4}2\pi\delta(k^2)\Theta(\omega),
  \label{dPhidef}
\end{equation}
and we have introduced the constants $\{N_n\}$, which collect
numerical factors and coupling constants. In words,
eq.~(\ref{phiFourier}) represents the spacetime field as an on-shell
inverse Fourier transform of the momentum-space amplitude. It was
derived in ref.~\cite{Monteiro:2020plf} using the so-called KMOC
formalism for obtaining classical observables from quantum field
theory\footnote{To convert to the notation of
ref.~\cite{Monteiro:2020plf}, one must write $\omega^n
\lambda_{A_1}\ldots \lambda_{A_{2n}}\equiv |k\rangle|k\rangle\ldots
|k\rangle$ in eq.~(\ref{phiFourier}), such that there are $2n$ factors
of the spinor $|k\rangle$. Furthermore, retarded boundary conditions
for the radiated field are implicit in eq.~(\ref{phiFourier}), which
is equivalent to slightly deforming the radiated energy according to
$k^0\rightarrow k^0+i\varepsilon$, $\varepsilon>0$.}. A similar
conclusion was presented in ref.~\cite{Guevara:2021yud}, using
different but related arguments. However, whereas
ref.~\cite{Monteiro:2020plf} examined the position-space implications
of eq.~(\ref{phiFourier}) directly by carrying out the inverse Fourier
transform in one go, ref.~\cite{Guevara:2021yud} split this into two
stages, according to the scheme of figure~\ref{fig:guevara}. To see
how this works in the present context (i.e. for eq.~(\ref{phiFourier})
taken from ref.~\cite{Monteiro:2020plf}), we may recast
eq.~(\ref{dPhidef}) so as to involve the variables appearing in
eq.~(\ref{kAA'}). To do this, we may equate $k_{AA'}$ (obtained from
eq.~(\ref{Vspinor})) with the right-hand side of eq.~(\ref{kAA'}), and
solve for the components $k_a$. In doing so, one may use
eq.~(\ref{lambdaparam}) and also the fact that nullity of the
reference vector $q_a$ implies
\begin{displaymath}
  q_{AA'}=q_A\tilde{q}_{A'},
\end{displaymath}
for some spinors $q_A$, $\tilde{q}_{A'}$. One finds
\begin{align}
  k_0&=\frac12\left[\xi \left(q_0\tilde{q}_{0'}+q_1\tilde{q}_{1'}\right)
    +\omega(1+z\tilde{z})\right];\notag\\
  k_1&=\frac12\left[\xi\left(q_0 \tilde{q}_{1'}-q_1\tilde{q}_{0'}\right)
    +\omega(z-\tilde{z})\right];\notag\\
  k_2&=\frac12 \left[\xi\left(q_0\tilde{q}_{0'}-q_1\tilde{q}_{1'}\right)
    +\omega(1-z\tilde{z})\right];\notag\\
  k_3&=\frac12\left[\xi\left(q_0\tilde{q}_{1'}+q_1\tilde{q}_{0'}
    +\omega(z+\tilde{z})\right)\right].
\label{ksol}
\end{align}
The Jacobian is given by \cite{Cachazo:2004kj}
\[
J = \frac{i\omega^2}{4\nu}\lambda^Aq_A\tilde{\lambda}^{A'}\tilde{q}_{A'},~~~~~\nu = \begin{cases}
	1~~~~~&\text{in (1,3) signature}\\
	i~~~~~&\text{in (2,2) signature},
\end{cases}
\]
from which one finds 
\begin{align}
  d\Phi(k)&=\frac{idz \,d\tilde{z}\,d\omega \,
  d\xi}{4\nu(2\pi)^3}\omega^2 (q_1-q_0\tilde{z})
  (\tilde{q}_{1'}-\tilde{q}_{0'}z)\, \delta[\xi\omega(q_1-q_0\tilde{z})
    (\tilde{q}_1-\tilde{q}_0 z)]\notag\\
  &=\frac{idz \,d\tilde{z}\,d\omega \,
  d\xi}{4\nu(2\pi)^3}\omega\delta(\xi).
  \label{dPhitrans}
\end{align}
Note that the on-shell delta function simply becomes the requirement
that $\xi=0$, as expected from the parametrisation of
eq.~(\ref{kAA'}). Substituting eq.~(\ref{dPhitrans}) into
eq.~(\ref{phiFourier}), one obtains\footnote{Carrying through the
$i\varepsilon$ prescription for the retarded boundary conditions in
eq.~(\ref{dPhitrans}) amounts to the deformation $\xi\rightarrow
\xi-i\omega \varepsilon$ in the first line of eq.~(\ref{int1}), which
guarantees convergence of the energy integral in the second line.}
\begin{align}
  \phi_{A_1\ldots A_{2n}}&=\frac{N_n}
      {4(2\pi)^2}\int dz \,d\tilde{z}\,d\omega \,
  d\xi\,
  \delta(\xi) \,\delta(2p\cdot k)\,\omega^{n+1}\,\lambda_{A_1}\ldots
  \lambda_{A_{2n}} e^{-\frac{i\omega}{2}\lambda_A\tilde{\lambda}_{A'}x^{AA'}}
  e^{-i\xi q\cdot x}{\cal A}_+(k)\notag\\
  &=
  \frac{N_n}{4(2\pi)^2}\int dz d\tilde{z}\,d\omega 
  \,\delta(2p\cdot k)\,\omega^{n+1}\,\lambda_{A_1}\ldots
  \lambda_{A_{2n}} e^{-\frac{i\omega}{2}\lambda_A\tilde{\lambda}_{A'}x^{AA'}}
  {\cal A}_+(k),
  \label{int1}
\end{align}
where we have used eq.~(\ref{VdotW}), and eliminated the on-shell
delta function in the second line. For brevity, we have also left
implicit the overall real part from eq.~(\ref{phiFourier}), and will
continue to do so in what follows. Regarding the remaining integral
measure, we may write this in a spinor-invariant form as
follows. First, from eq.~(\ref{lambdaparam}), we may rewrite
\begin{displaymath}
  \int dz\rightarrow \oint \lambda_E d\lambda^E,
\end{displaymath}
where the latter is the conventional measure on the Riemann sphere
associated with $\lambda_E$. Note that, from the parametrisation of
eq.~(\ref{lambdaparam}), the integral over $z$ is in the complex plane
one obtains by stereographic projection. Rewritten in terms of
$\lambda_E$, this will become a closed contour integral on the Riemann
sphere itself. Next, we may define
\begin{equation}
  \tilde{\xi}_{A'}=\omega \tilde{\lambda}_{A'}=\omega(1,\tilde{z}),
  \label{mutildedef}
\end{equation}
such that one has
\begin{equation}
  d\omega d\tilde{z}=\frac{1}{\omega}d\tilde{\xi}_{0'}d\tilde{\xi}_{1'}
  \equiv \frac{1}{\omega}d^2\tilde{\xi}.
  \label{measuremu}
\end{equation}
Equation~(\ref{int1}) then becomes
\begin{equation}
  \phi_{A_1\ldots A_{2n}}=\frac{1}{2\pi i}\oint \lambda_E d\lambda^E\,
  \lambda_{A_1}\ldots \lambda_{A_{2n}}\,\rho_x
  \left[\mathfrak{M}_+(Z^\alpha)\right],
\label{int2}
\end{equation}
where we have defined
\begin{equation}
  \mathfrak{M}_+=\frac{i N_n}{4(2\pi)^2}\int d^2\tilde{\xi}
  e^{-\frac{i}{2}\lambda_A \tilde{\xi}_{A'} x^{AA'}}
  \delta(2p\cdot k)\omega^{n}
  {\cal A}_+(k).
  \label{gothicM}
\end{equation}
We may recognise eq.~(\ref{int2}) as precisely the Penrose transform
of eq.~(\ref{Penrose}), where the spinor $\lambda_A$ entering the
spinorial decomposition of the radiation momentum of
eq.~(\ref{lambdaparam}) forms half of the twistor components defined
in eq.~(\ref{dualtwistor}). Using the incidence relation of
eq.~(\ref{incidence}), we may then recognise the combination
$\lambda_A x^{AA'}$ appearing in the exponent in eq.~(\ref{gothicM})
as the remaining half of the twistor $\mu^{A'}$. Hence, one has
\begin{equation}
  \mathfrak{M}_+(Z^\alpha)=\int d^2\tilde{\xi}
  e^{-\frac{i}{2}\tilde{\xi}_{A'} \mu^{A'}} A(k),\quad
  A(k)=\frac{i N_n}{4(2\pi)^2}
  \delta(2p\cdot k)\omega^{n}
  {\cal A}_+(k).
  \label{gothicM2}
\end{equation}
The object on the left-hand side depends on the spinors $\lambda_A$
and $\mu^{A'}$, and hence the single twistor argument
$Z^\alpha$. We note that the little group properties of the amplitude, i.e. that it transforms as ${\cal A} \rightarrow t^{-2n}{\cal A}$ under a little group transformation of the massless leg, ensures that $\mathfrak{M}$ transforms with the required homogeneity of $-2n - 2$ , since the measure transforms as $d^2\tilde{\xi} \rightarrow t^{-2}d^2\tilde{\xi}$. The integral transform appearing in
eq.~(\ref{gothicM2}) takes a certain dressed momentum-space amplitude,
and maps it into a quantity in twistor space. We will call this the
``half transform'' as in figure~\ref{fig:guevara} given that, as
pointed out in ref.~\cite{Guevara:2021yud}, eq.~(\ref{gothicM2}) is
related to the well-known ``half Fourier transform'' of
refs.~\cite{Witten:2003nn,ArkaniHamed:2009si,Mason:2009sa} that takes
momentum-space amplitudes into twistor space.  Here and in
ref.~\cite{Guevara:2021yud}, however, one integrates only over
manifestly positive energies $\omega>0$ such that the half transform,
considered as an integral in $\omega$, is equivalent to a Laplace
transform, as we will see in explicit examples.\\

The above results indeed realise the scheme of
figure~\ref{fig:guevara}: eq.~(\ref{gothicM2}) is the half transform
mapping momentum-space quantities into twistor space. Subsequently,
eq.~(\ref{int2}) is the Penrose transform that takes the quantity
$\mathfrak{M}(Z^\alpha)$ in twistor space, and associates it with a
classical spacetime field. Thus, in a well-defined sense, twistor
space sits ``in between'' momentum and position space, allowing us to
address conceptual questions regarding the double copy.\\

\subsection{Cohomology representatives from half-transformed amplitudes}
\label{sec:ampreps}

As discussed in section~\ref{sec:review}, an open problem in the
twistor double copy is to make sense of the product of cohomology
representatives occuring in eq.~(\ref{fgrav}). Whilst various ideas
for choosing representatives have occurred in recent
literature~\cite{Chacon:2021wbr,Adamo:2021dfg,Chacon:2021lox}, it is
not clear that any of these correspond to the original \u{C}ech double
copy presented in
refs.~\cite{White:2020sfn,Chacon:2021wbr}. Furthermore, it would be
reassuring to know that any incarnation of the twistor double copy can
be shown to be equivalent to the original BCJ double copy for
scattering amplitudes~\cite{Bern:2010ue,Bern:2010yg}, which would
immediately put the twistor approach on a much firmer footing. In
fact, the scheme of figure~\ref{fig:guevara} allows us to do just
this. First, note that the half transform of eq.~(\ref{gothicM2})
relates a given momentum-space amplitudes to a specific (unambiguous)
cohomology representative in twistor space. From a double copy point
of view, there is then a natural choice of representative for a given
classical solution in position space, namely that which is picked out
by a momentum-space amplitude. For certain amplitudes relating to
known static solutions in position space, we show that the cohomology
representatives in twistor space are {\it precisely} those entering
the \u{C}ech double copy of eq.~(\ref{fgrav}). This in turn implies
that for these solutions, the BCJ double copy for
amplitudes~\cite{Bern:2010ue,Bern:2010yg}, the twistor double copy of
refs.~\cite{White:2020sfn,Chacon:2021wbr}, and the Weyl double copy of
ref.~\cite{Luna:2018dpt}, amount to the same thing. Whilst the
connection between the Weyl double copy and three-point amplitudes in
momentum space was already noted in ref.~\cite{Monteiro:2020plf}, the
linking of both of these to an intermediate twistor space is both new
and useful, as we will see later on. \\

Let us now find the cohomology representatives in twistor space
corresponding to given amplitudes. If we take a spinless static source with
\begin{equation}
	p^\mu=Mu^\mu,
	\label{pdef}
\end{equation}
where $u^\mu$ is the 4-velocity, then the delta function appearing in
eq.~(\ref{gothicM2}) becomes, when translated using
eq.~(\ref{lambdaparam}),
\begin{equation}
	\delta(2p\cdot k)=\delta (M u^{AA'}\lambda_A\tilde{\xi}_{A'}),
	\label{deltafunction}
\end{equation}
which implies
\begin{equation}
	\tilde{\xi}_{A'} \propto \omega {u^A}_{A'}\lambda_A.
	\label{xidelta}
\end{equation}
In the rest frame where $u_a = (1,\textbf{0})$, we find that
\[
\omega u^A_{~A'}\lambda_A = z\tilde{\xi}_{A'}.
\]
We recall the fact that we have the freedom to perform a scaling of the form $\lambda \rightarrow t \lambda$ and $\tilde{\lambda} \rightarrow t^{-1} \tilde{\lambda}$, and that the curvature spinor is invariant under such scaling. We choose to scale
\[
\lambda \rightarrow \frac{1}{\sqrt{z}} \lambda,~~~~~\tilde{\lambda} \rightarrow -\frac{1}{\sqrt{-\tilde{z}}} \tilde{\lambda},
\]
using the fact that eq. \eqref{deltafunction} fixes $z\tilde{z} = -1$. By fixing this symmetry, eq. \eqref{xidelta} becomes an equality
\[
\tilde{\xi}_{A'} = \omega {u^A}_{A'}\lambda_A.
\label{xideltaequality}
\]

Equation~(\ref{gothicM2}) can be expressed as\footnote{We have removed the
helicity subscript in eq.~(\ref{Mscal}), given that this is not
relevant for the scalar case.}
\begin{equation}
    \mathfrak{M}(Z^\alpha)=\int_0^\infty d\omega
    \exp\left[{-\frac{\omega}{2}{u^A}_{A'}\mu^{A'}\lambda_A }\right] {\cal A}(k)
    \label{Mscal}
\end{equation}
where, following ref.~\cite{Guevara:2021yud}, we have rotated
$x^\mu\rightarrow i x^\mu$. The latter replacement merely corresponds
to reversing the (2,2) signature of our spacetime metric, and avoids
proliferation of factors of $i$ in what follows. The amplitude for a
scalar is simply a coupling, and so one may straightforwardly carry
out the $\omega$ integral to give
\begin{equation}
  \mathfrak{M}\propto\frac{1}{{u^A}_{A'}\mu^{A'}\lambda_A}.
  \label{Mscal2}
\end{equation}
For a static particle we must have $u_a=(1,\vec{0})$, and it then
follows from eq.~(\ref{Vspinor}) that
\begin{equation}
  {u^A}_{A'}=\epsilon^{BA}u_{BA'}=\left(\begin{array}{cc}0 & -1\\1 & 0
  \end{array}\right).
  \label{uAA'static}
\end{equation}
Then eq.~(\ref{Mscal2}) may be rewritten as
\begin{equation}
  \mathfrak{M}\propto \frac{1}{Q_{\alpha\beta}Z^\alpha Z^\beta},\quad
  Q_{\alpha\beta} = \left(\begin{array}{cc}0 & {u^{A}}_{B'}\\
  	{u^{B}}_{A'} & 0\end{array}\right) = \left(\begin{array}{cccc}0 & 0 & 0 & -1\\
    0 & 0 & 1 & 0 \\ 0 & 1 & 0 & 0\\
    -1 & 0 & 0 & 0\end{array}\right).
  \label{Mscal3}
\end{equation}
We have thus obtained a cohomology representative in twistor space for
a static scalar field, and we can immediately note that this has
precisely the form required by the \u{C}ech form of the twistor double
copy i.e. eq.~(\ref{f-4Q}). Furthermore, the explicit form of
$Q_{\alpha\beta}$ is indeed that used to obtain the zeroth copy of the
Schwarzschild black hole in refs.~\cite{White:2020sfn,Chacon:2021wbr}
(see also refs.~\cite{Quadrille,Hughston:1979tq} for the original
context in which this quadratic form was presented). \\

A generic amplitude of two scalars with mass $M$ and a positive-helicity spin-$n$ radiated field is given by
\[
\mathcal{A}^{(n)}_+(k) = g_n M^n X^n,~~~~~~X = \sqrt{2}u\cdot\epsilon_+(k).
\]
We will focus on the electromagnetic and gravitational cases, where
the relevant amplitudes are given by taking $g_1 = -\sqrt{2}Q$ and $g_2 = -\frac{\kappa}{2}$ (see
e.g. ref.~\cite{Monteiro:2020plf}) and
\begin{equation}
  {\cal A}^{\rm EM}_+=-\sqrt{2}MQX,
  \quad {\cal A}_+^{\rm grav.}=-\frac{\kappa}{2} M^2 X^2.
  \label{amps}
\end{equation}
The fact that the gravity amplitude in eq.~(\ref{amps}) is related to
the square of the electromagnetic case is a manifestation of the BCJ
double copy for amplitudes. At this stage it is useful to introduce
the bispinor
\[\label{Tspinor}
T_{AA'} = V_a\sigma^{a}_{AA'},
\]
where the vector $V_a$ is a fixed constant vector with $V^2 = \pm 1$ that points in only one direction. This bispinor has an important property, namely that
\[
T_{AA'}T^{A'B} = \pm\delta^{B}_A.
\]
To see this, we use the Clifford algebra
\[
T_{AA'}T^{A'B} = V_a\sigma^{a}_{AA'}V^b\bar{\sigma}_{b}^{A'B} = \frac12V_aV_b\left[\sigma^{a}_{AA'}\bar{\sigma}^{b,A'B} + \sigma^{b}_{AA'}\bar{\sigma}^{a,A'B}\right] = V^2\delta^B_A = \pm \delta^B_A, 
\]
where the bar denotes Infeld-van-der-Waerden symbols acting on
conjugate spinors. For static solutions, $u_a = (1,\textbf{0})$ and
the bispinor $u^{AA'}$ is exactly of the form of
eq. \eqref{Tspinor}. This means we can recast the delta function
constraint, in the static case, to be
\[\label{staticDF}
u^{BA'}\tilde{\xi}_{A'}= \omega {u^A}_{A'}u^{A'B}\lambda_A = \omega\lambda_B.
\]
Using this, the delta function constrains the three-particle amplitudes to be simple constant factors. One way to see this is to use the explicit spinor form of the polarisation
vector\footnote{In the conventional spinor helicity notation of
	ref.~\cite{Monteiro:2020plf}, eq.~(\ref{epsilon+}) reads
	$\epsilon_+^a=\frac{[k|\tilde{\sigma}^a|q\rangle}{\sqrt{2}\langle k
		q\rangle}$, where $\tilde{\sigma}^a$ denotes an
	Infeld-van-der-Waerden symbol with upstairs spinor indices.}
\begin{equation}
	\epsilon^a_+=\frac{1}{\sqrt{2}}\frac{(\sigma^a)^{AA'}q_A
		\tilde{\lambda}_{A'}}{\lambda^A q_A},
	\label{epsilon+}
\end{equation}
where $q_A$ is a so-called {\it reference spinor}, corresponding to a
null reference vector in the tensorial language. The form of
$u_a=(1,\vec{0})$ for a static solution implies
\begin{equation}
	X = \sqrt{2}u\cdot \epsilon_+=\frac{u^{AA'}q_A
		\tilde{\lambda}_{A'}}{\lambda^A q_A}= \frac{1}{\omega}\frac{u^{AA'}q_A
		\tilde{\xi}_{A'}}{\lambda^A q_A} = 1,
	\label{epsilon+2}
\end{equation}
where we have again used \eqref{staticDF}. We see then that the $X$-factor is set to unity and the spin-$n$ amplitude, under the integral of the Laplace transform, is simply proportional to some coupling times a mass. Using this fact, and carrying out similar steps to the scalar case,
we find that the spin-$n$ version of eq.~(\ref{gothicM2}) satisfies
\begin{align}
  \mathfrak{M}_{n,+}(Z^\alpha)&\propto\int_0^\infty d\omega \,\omega^n
  \exp\left[-\frac{\omega}{2}{u^{A}}_{A'}\mu^{A'}\lambda_A\right]\notag\\
  &\propto \frac{1}{(Q_{\alpha\beta}Z^\alpha Z^\beta)^{n+1}},
  \label{Mn+}
\end{align}
where $Q_{\alpha\beta}$ is given by eq.~(\ref{Mscal3}) as before. For
spinless static solutions -- corresponding physically to the
Schwarzschild black hole and its single / zeroth copies -- we have
thus reproduced the cohomology representatives of eq.~(\ref{f-6Q},
\ref{f-4Q}). \\

It is straightforward to generalise the above arguments to other
classical solutions, that are related to three-point amplitudes
according to eq.~(\ref{phiFourier}). As argued in
refs.~\cite{Emond:2020lwi,Monteiro:2020plf,Monteiro:2021ztt}, for example, one may
modify the 3-point amplitudes for a spinless static particle to
include both the effects of rotation, and also a dual charge e.g. a
NUT charge in gravity, corresponding to a magnetic monopole in gauge
theory~\cite{Luna:2015paa}. Furthermore, this modification is
remarkably simple: one simply replaces the three-point amplitude for
helicity $\eta$ according to\footnote{For the scalar field, one
chooses the sign of $\eta$ according to whether one is taking the
zeroth copy of the self-dual or anti-self-dual electromagnetic field
strength spinor.}
\begin{equation}
  {\cal A}_\eta\rightarrow e^{\eta (ik\cdot a+\theta)}{\cal A}_\eta,
  \label{KerrNUT}
\end{equation}
where $\theta$ is related to the NUT charge, and $a^\mu$ is the
classical spin vector. In twistor space, this has the effect of simply
multiplying each representative by $e^{\theta}$, whilst simultaneously
replacing $x\rightarrow x-a$, so that we get
\begin{align}
  \mathfrak{M}_{n,+}&\propto\frac{e^\theta}{({u^A}_{A'}\mu^{A'}\lambda_A
    -{u^{(A}}_{A'} a^{B)A'}\lambda_A\lambda_B)^{n+1}}\notag\\
  &\propto\frac{1}{(Q_{\alpha\beta}Z^\alpha Z^\beta)^{n+1}},
  \label{MKerrNUT}
\end{align}
where we now have
\begin{equation}
  Q_{\alpha\beta}=e^{-\theta}\left(\begin{array}{cc}0 & {u^{A}}_{B'}\\
    {u^{B}}_{A'} & 2\mu^{AB}\end{array}
  \right),\quad
  \mu^{AB}=-{u^{(A}}_{A'} a^{B)A'}.
  \label{kinematic}
\end{equation}
Up to our overall normalisation, the result for $Q_{\alpha\beta}$ (in
the case $\theta=0$) is precisely the so-called {\it kinematic
  twistor} that encodes the (angular) momentum of a spinning
particle~\cite{Penrose:1972ia,Penrose:1986ca}\footnote{Note that the
lower-right components in eq.~(\ref{kinematic}) have an accompanying
factor of $i$ in refs.~\cite{Penrose:1972ia,Penrose:1986ca}, owing to
the choice of Lorentzian rather than (2,2) signature.}. Again, we find
that the cohomology representatives picked out by momentum-space
three-point amplitudes are precisely the quadratic forms required by
eqs.~(\ref{f-6Q}, \ref{f-4Q}). \\

Above, we remarked that the half transform from momentum to twistor
space assumes the form of a Laplace transform, and we can see this
directly in our explicit examples. First, note that the right-hand
side of eq.~(\ref{Mn+}) may be written as
\begin{equation}
  {\cal L}[\omega^n](U)\equiv \int_0^\infty d\omega \,\omega^n
  e^{-\omega U},\quad U=\frac12 {u^A}_{A'}\mu^{A'}\lambda_A,
  \label{Laplace1}
\end{equation}
which is a manifest Laplace transform in the energy $\omega$, with $U$
playing the role of the conjugate variable. Likewise, the shift
$x\rightarrow x-a$ above eq.~(\ref{MKerrNUT}) amounts to the
replacement
\begin{equation}
  U\rightarrow U-V,\quad V=\mu^{AB}\lambda_A\lambda_B.
  \label{Ushift}
\end{equation}
This is commonly referred to as the {\it frequency-shifting} property
of Laplace transforms:
\begin{equation}
  {\cal L}[e^{\omega V} f(\omega)](U)={\cal L}[f(\omega)](U-V),
  \label{frequencyshift}
\end{equation}
which we collect for later use.\\

Let us take stock of what has happened. The twistor double copy of
refs.~\cite{White:2020sfn,Chacon:2021wbr} gave a way to ``derive'' the
Weyl double copy for type-D vacuum solutions in position space, but
suffered from the conceptual puzzle of how to multiply together the
relevant quantities in twistor space or, in other words, how to choose
appropriate cohomology representatives so that a simple product
structure is obtained. Procedures for achieving the latter were
discussed in refs.~\cite{Adamo:2021dfg,Chacon:2021lox}, but it remains
unclear whether or not these are equivalent. Furthermore, none of them
reproduces the original choice of representatives in the original
twistor double copy of refs.~\cite{White:2020sfn,Chacon:2021wbr}. In
this section, we have used the methods of
refs.~\cite{Monteiro:2020plf,Guevara:2021yud} to show that, for those
stationary fields which can be obtained from three-point amplitudes in
momentum space, then the Weyl double copy~\cite{Luna:2018dpt}, the
twistor double copy~\cite{White:2020sfn,Chacon:2021wbr}, and the BCJ
double copy for scattering amplitudes~\cite{Bern:2010ue,Bern:2010yg}
{\it are completely equivalent}. They are related by integral
transforms according to the scheme of figure~\ref{fig:guevara}, and
our findings are significant in that they immediately put the twistor
double copy on a much firmer footing. Furthermore, they suggest how it
may be extended (e.g. by half-transforming considering more
complicated amplitudes in momentum space). For the remainder of this
paper, we discuss some of the conceptual implications of
figure~\ref{fig:guevara}, specifically regarding locality of the
Weyl double copy.

\section{Why is the type-D Weyl double copy local in position space?}
\label{sec:simple}

In the previous section, we have shown that the twistor version of the
type-D Weyl double copy can be straightforwardly obtained by
half-transforming three-point amplitudes from momentum space. As well
as justifying the use of twistor methods in studying the double copy,
this allows us also to examine certain conceptual questions regarding
exact classical double copies. As remarked above, the fact that the
exact position-space double copies are possible is puzzling, given
that the traditional BCJ double copy for scattering ampitudes is set
up in momentum space. In fact, the construction of
figure~\ref{fig:guevara} extends this interesting phenomenon, given
that the momentum-, twistor- and position-space double copies are all
related by integral transforms. Each one of these transforms is
non-local, and yet the type-D Weyl double copy is manifestly local in
all three spaces, involving products of quantities evaluated at the
same point. How can this possibly be true?

\subsection{Locality in twistor space}
\label{sec:whylocal?}

The first step in answering these questions is to consider the half
transform of momentum-space amplitudes into twistor space. The BCJ
double copy in momentum space is multiplicative, in that the
amplitudes entering eq.~(\ref{gothicM2}) for different theories are
related by
\begin{equation}
  {\cal A}_+^{\rm grav.}=\frac{{\cal A}_{+}^{\rm EM}{\cal A}_{+}^{\rm EM}}
  {{\cal A}^{\rm scal.}}.
  \label{ampcopy}
\end{equation}
This multiplicative structure survives upon including the energy
dependence from eq.~(\ref{gothicM2}) i.e. the overall power of
$\omega$. To go to twistor space in each theory, we must use a Laplace
transform in the energy, as discussed in eqs.~(\ref{Laplace1},
\ref{frequencyshift}). But this then creates the puzzle of why the
twistor-space representatives are related by the simple product of
eq.~(\ref{fgrav}), rather than the convolution one expects upon taking
the Laplace transform of a product.\\

The resolution of this puzzle lies in the very particular form of
three-point amplitudes in momentum space. The product in twistor space
will emerge provided that the amplitudes (and related energy factors)
in momentum space are such that their convolution is equivalent to a
product of similar functions. Clearly this is not true for most
functions, but it does happen to be true for pure power-like functions
of energy. Considering
\begin{equation}
  f(\omega)=\omega^\alpha,\quad g(\omega)=\omega^\beta,
  \label{fgomega}
\end{equation}
one has
\begin{equation}
  f(\omega)\star g(\omega)\equiv \int_0^\omega du\, f(u)g(\omega-u)=
  \frac{\Gamma(\alpha+1)\Gamma(\beta+1)}{\Gamma(\alpha+\beta+2)}
  \omega^{\alpha+\beta+1}.
  \label{fgconvolve}
\end{equation}
That is, the convolution of two power-like functions is also a
power-like function, up to an overall numerical factor. Denoting the
three-point amplitude for a spin-$n$ emission together with its
accompanying energy factor by
\begin{equation}
  \tilde{{\cal A}}^{(n)}=\omega^n \tilde{A}^{(n)}_+,
  \label{Andef}
\end{equation}
we may then write the gravity case of eq.~(\ref{Mn+}) as
\begin{align}
  \mathfrak{M}_{2,+}(Z^\alpha)&\propto \int_0^\infty e^{-\omega U}
  \tilde{{\cal A}}^{(1)}\tilde{{\cal A}}^{(1)}
        [\tilde{{\cal A}}^{(0)}]^{-1}\notag\\
        &\propto \int_0^\infty e^{-\omega U}
  \tilde{{\cal A}}^{(1)}\star\tilde{{\cal A}}^{(1)}
        \star[\tilde{{\cal A}}^{(0)}]^{-1},
        \label{momspacecon}
\end{align}
where $U$ has been defined in eq.~(\ref{Laplace1}), and the second
line follows from eq.~(\ref{fgconvolve}) and associativity of the
convolution. The convolution theorem then implies that the
gravitational twistor space representative is given by
\begin{equation}
  \mathfrak{M}_{2,+}(Z^\alpha)\propto
  \frac{{\cal L}[\tilde{{\cal A}}^{(1)}]{\cal L}[\tilde{{\cal A}}^{(1)}]}
       {{\cal L}[\tilde{{\cal A}}^{(0)}]}\propto
       \frac{\mathfrak{M}_{1,+}(Z^\alpha)\,\mathfrak{M}_{1,+}(Z^\alpha)}
            {\mathfrak{M}_{0,+}(Z^\alpha)},
       \label{M2prod}
\end{equation}
which is precisely the twistor-space product of
eq.~(\ref{fgrav}). Note that the power-like functions in
eqs.~(\ref{fgomega}, \ref{fgconvolve}) are not the only possibilities
that lead to a product in twistor space. One may also perform a {\it
  frequency shift} in the Laplace transform, according to
eq.~(\ref{frequencyshift}). Combining the latter with the convolution
theorem, it is easy to prove that the frequency shift operation
commutes with a convolution:
\begin{equation}
  {\cal L}[e^{\omega V}f(\omega)\star g(\omega)](U)
  ={\cal L}\left[\left(e^{\omega V}f(\omega)\right)\star
    \left(e^{\omega V}g(\omega)\right)\right](U).
  \label{freqshift2}
\end{equation}
The Kerr-Taub-NUT twistor representative of eq.~(\ref{MKerrNUT}) is
obtained by frequency shifting eq.~(\ref{momspacecon}) (as well as
multiplying by a constant factor):
\begin{equation}
  \mathfrak{M}_{2,+}=e^{\theta}{\cal L}
           [ e^{\omega V} \tilde{{\cal A}}^{(1)}\star\tilde{{\cal A}}^{(1)}
             \star(\tilde{{\cal A}}^{(0)})^{-1}].
           \label{KTN1}
\end{equation}
Using eq.~(\ref{freqshift2}), this is equivalent to shifting each
amplitude combination {\it before} taking the convolution:
\begin{equation}
  \mathfrak{M}_{2,+}={\cal L}
           [  (e^{\omega V+\theta}\tilde{{\cal A}}^{(1)})\star
             (e^{\omega V+\theta}\tilde{{\cal A}}^{(1)})
             \star(e^{\omega V+\theta}\tilde{{\cal A}}^{(0)})^{-1}],
           \label{KTN2}
\end{equation}

In twistor space, this means that the product form of
eq.~(\ref{M2prod}) remains the same, even for the full Kerr-Taub-NUT
solution. The procedure of eq.~(\ref{KerrNUT}), that relates the
amplitudes for the Kerr-Taub-NUT solution to those generating pure
Schwarzschild, is a momentum-space counterpart of the well-known {\it
  Newman-Janis shift} for the corresponding classical
fields~\cite{Newman:1965tw}. Here we see a novel interpretation of
this shift, namely that it acts as a frequency shift in the energy
Laplace transform of the momentum-space amplitude, whose consequence
is to ensure locality of the double copy in twistor space!\\

In summary, the twistor double copy for type-D solutions involves a
local product of cohomology representatives because: (i) these
cohomology representatives can be obtained as a half transform
(equivalent to a Laplace transform in energy) of momentum-space
amplitudes; (ii) the form of the amplitudes is precisely such that
their product in momentum space is equivalent to a convolution. Next,
let us consider why locality in twistor space implies locality in
position space.

\subsection{Locality in position space}
\label{sec:localposition}

As we have seen above, locality of the double copy in momentum space
implies locality in twistor space given the specific form of
three-point amplitudes, and also the mathematical properties of the
half transform, which is equivalent to a Laplace transform in
energy. The map between twistor space and position space also involves
some non-locality, although the nature of the integral transform is
different, as is clear from figure~\ref{fig:guevara}. Thus, a
different argument is needed to explain why the type-D Weyl double
copy is local in position space, given locality in twistor space.\\

First, let us remind ourselves that all of the \u{C}ech cohomology
representatives for type-D solutions involve inverse powers of a
quadratic form in the twistor variable $Z^\alpha$. This implies the
presence of two poles in twistor space, that will appear on the
Riemann sphere $X$ corresponding to each spacetime point $x^{AA'}$
once the appropriate incidence relation of eq.~(\ref{incidence}) is
imposed. The Penrose transform is a contour integral that will pick
out the residue of one of these poles, which amounts to the vanishing
of a twistor function. This has a nice interpretation in position
space, due to the following result known as the {\it Kerr Theorem}
(see e.g. ref.~\cite{Penrose:1986ca} for an extended discussion, and
refs.~\cite{Elor:2020nqe,Farnsworth:2021wvs} for a complementary
application of the Kerr theorem to understanding the classical double
copy):\\

{\it Given a holomorphic, homogeneous twistor function
  $\chi(Z^\alpha)$, the requirement $\chi(Z^\alpha)=0$ defines a
  null shear-free geodesic vector field in Minkowski space.}\\

\begin{figure}
  \begin{center}
    \scalebox{0.6}{\includegraphics{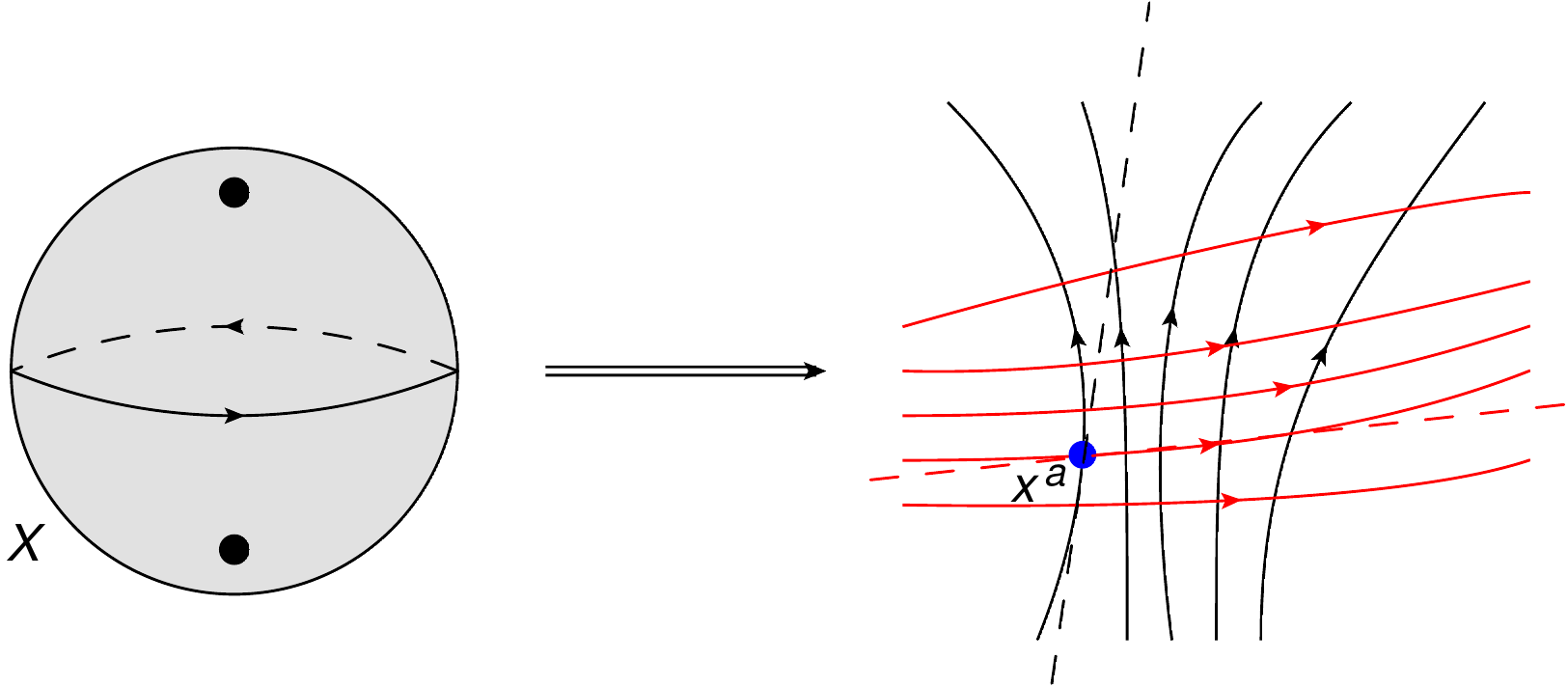}}
    \caption{The poles of a twistor ``function'' $f(Z^\alpha)$ define
      shear-free null geodesic congruences in spacetime. For a given
      spacetime point, carrying out the Penrose transform on the
      Riemann sphere $X$ in $\mathbb{PT}$ picks out the null
      directions at a single spacetime point (shown in blue). One thus
      obtains a local product of principal spinors in spacetime, even
      though the twistor product implies a non-local statement in
      position space.}
    \label{fig:twistor_congruences}
  \end{center}
\end{figure}

The two poles in the inverse quadratic form for type-D solutions thus
imply the presence of two null shear-free geodesic vector fields in
spacetime, which is indeed a characteristic feature of type D
solutions. Note that the Kerr theorem applies for all spacetime points
simultaneously, given that it applies to the full twistor function
entering the Penrose transform, {\it before} restriction to a given
spacetime point $x^{AA'}$. This situation is depicted in
figure~\ref{fig:twistor_congruences}, where on the right-hand side we
draw two null shear-free vector fields in spacetime. In twistor space,
restriction to a given spacetime point leads to a particular Riemann
sphere $X$, corresponding to the blue point $x^a$ on the right-hand
side of the figure. The poles of the general twistor function, upon
restriction to this Riemann sphere $X$, correspond to fixed points in
projective twistor space $\mathbb{PT}$. As discussed in
section~\ref{sec:review}, points in $\mathbb{PT}$ correspond to null
geodesics in Minkowski space. The latter will be tangent to the null
shear-free vector fields generated by the general twistor function
corresponding to a given pole, as shown on the right-hand side of the
figure. These null directions are in one-to-one correspondence with
the principal spinors $\alpha_A$, $\beta_A$ of the spacetime field at
the point $x^a$.  \\

The twistor double copy for type-D solutions states that one may
combine the representatives of scalar and gauge fields of
eq.~(\ref{f-4Q}) in order to obtain the gravity representative of
eq.~(\ref{f-6Q}). This does not change the location of the poles in
twistor space, such that all of the spacetime fields entering the
correspondence have the {\it same} pair of null shear-free geodesic
vector fields associated with them. In the spinor language, this means
that the spacetime fields have the same pair of principal spinors
$\alpha_A(x)$ and $\beta_A(x)$, such that only their multiplicity
differs between theories. The type-D Weyl double copy of
eq.~(\ref{WeylDC}) then simply amounts to the statement that the
multiplicity of the principal spinors of a gravity solution can be
simply obtained by appropriately combining the principal spinors of
gauge and scalar fields. This is both a non-local and a local
statement. It is non-local in that it is a statement about {\it
  directions}, and thus entire null geodesics associated with a given
spacetime point. This is precisely the non-locality one expects upon
transforming a local product in twistor space into a space-time
statement. However, the Weyl double copy is local in that it refers to
the principal spinors at a given spacetime point, which are associated
with the null tangent directions on the right-hand side of
figure~\ref{fig:twistor_congruences}. The null directions are
potentially different at all points in spacetime, and this is
reflected in the Weyl double copy by the fact that it applies
point-by-point in spacetime. \\

\subsection{Beyond type-D solutions}
\label{sec:beyond}

In the previous two sections, we have used the construction of
figure~\ref{fig:guevara} to argue why the type-D Weyl double copy is local in
position space. The intermediate twistor step is useful in this
regard, as it provides another layer of information that allows us to
visualise properties of spacetime fields geometrically. It also allows
to address when the simple properties embodied by the type-D Weyl
double copy might fail. We will exclude the case of non-vacuum
solutions, for which the techniques of this paper -- which heavily
rely on the Penrose transform for massless free fields -- do not
apply. Indeed, ref.~\cite{Monteiro:2021ztt} provided an example of a
non-vacuum solution which indeed does not have a local position-space
double copy. It was also not a pure gravity solution, due to a
non-zero dilaton field.\\

Assuming that the ``true'' double copy is in
momentum-space~\cite{Bern:2010ue,Bern:2010yg}, the results of
section~\ref{sec:whylocal?} tell us that locality in twistor space is
expected to fail for those amplitudes which are not proportional to
pure exponentials in the energy. In that case, the Laplace transform
to twistor space will involve non-power-like functions, such that a
more complex structure in twistor space is obtained, rather than the
simple product that is needed to reproduce the Weyl double copy. Thus,
local twistor-space double copies will not be obtained in general
beyond linearised level, given that the appropriate amplitudes in
momentum space will have non-trivial momentum dependence, including
poles in Mandelstam invariants. \\

Even if restricting to linearised level, one can ask if the simple
form of the Weyl double copy of eq.~(\ref{WeylDC}) is generic for
arbitrary (approximate) Petrov types. In
section~\ref{sec:localposition}, we used the Kerr theorem to argue
that the Penrose transform will automatically lead to a local
position-space double copy, for scalar, gauge and gravity fields that
share the same poles in twistor space (or principal spinors in
spacetime). Scrutiny of this argument reveals that it only depends on
there being a pair of common poles in the twistor representatives for
the fields. Allowing the multiplicity of these poles to be different
to that in the type-D case, one may obtain type III or N solutions,
which (from table~\ref{tab:Petrov}) all have at most a pair of distinct
spinors associated with them. Indeed, examples of such linearised
double copies have been given in
refs.~\cite{White:2020sfn,Chacon:2021wbr}. Where more than two poles
are present in twistor space, the situation is more
complicated. Performing the Penrose transform integral means that one
must take the residue of more than one pole in twistor space. The
resulting spacetime gravity field is given by a sum of Weyl
double-copy-like terms, such that the total principal spinors of the
gravity field are not necessarily easily related to those of the
constituent gauge and scalar fields~\cite{Chacon:2021wbr}. Thus, the
simple form of the Weyl double copy is indeed highly special, and is
not expected to be true in general for either non-linear fields, or
linear fields that have more than two principal null directions.

\section{Discussion}
\label{sec:discuss}

In this paper, we have explored the issue of why the well-known {\it
  Weyl double copy} relating fields in scalar, gauge and gravity
theories, is local in position space. This question arises given that
the original BCJ double copy for scattering
amplitudes~\cite{Bern:2010ue,Bern:2010yg} is local in momentum
space. Although recent works have shown that mathematical properties
of Fourier integrals from momentum to position space indeed imply
locality for some solutions~\cite{Monteiro:2021ztt}, we have here
sought a more underlying explanation. To this end, we have used the
ideas of ref.~\cite{Guevara:2021yud}, that say that one may split the
Fourier transform from momentum to position space into two steps. The
first takes three-point amplitudes into twistor space, which we have
shown leads to the twistor double copy of
refs.~\cite{White:2020sfn,Chacon:2021wbr}. The second step is a
Penrose transform, which produces the Weyl double copy for type D
solutions. By using known three-point amplitudes in momentum space, we
have shown that, for type-D solutions where relevant amplitudes are
known, the BCJ, twistor and Weyl double copies amount to the same
thing. \\

As a byproduct, our analysis resolves a lingering puzzle in the
twistor double copy, which involves products of ``functions'' in
twistor space. These functions should actually be interpreted as
representatives of cohomology classes, and one must then provide a
prescription for picking out special representatives. Various such
procedures have been given in the
literature~\cite{Adamo:2021dfg,Chacon:2021lox}, but none of them
obviously corresponds to the twistor double copy of
refs.~\cite{White:2020sfn,Chacon:2021wbr}. In this paper, we have
shown that the required representatives in twistor space are precisely
those picked out by three-point amplitudes! This observation may prove
very useful in extending the use of twistor methods in the double
copy. \\

The mechanism by which the type-D classical double copy inherits
locality in position space is interesting. First, the known amplitudes
corresponding to type-D fields in spacetime are such that their
convolution is equivalent to a product of similar functions. Thus, the
``half transform'' that takes amplitudes into twistor space implies
the presence of a local product in twistor space. This criterion will
fail beyond linearised level, confirming that local
position space double copies are highly special.\\

Secondly, the Penrose transform from twistor to position space has
both a local and a non-local character. The Weyl double copy is a
statement about {\it directions} (principal spinors), which are
non-local objects, in keeping with the fact that points in twistor
space are associated with null geodesics in position space. However,
the principal spinors of a field are different at different spacetime
points in general, so that the local information in the Weyl double
copy is simply that the principal spinors of gravity fields are
obtained from their gauge and scalar counterparts point-by-point in
spacetime. The simple nature of the Weyl double copy is restricted to
those solutions that share the same pair of poles in twistor space,
and hence have only two distinct principal spinors. For type II or
type I fields, this will no longer be true.\\

We hope that our results clarify the nature of exact position space
double copies, and confirm their rigour where applicable. We believe
our results also suggest that further use of twistor ideas will prove
fruitful in clarifying other aspects of the double copy
correspondence, which continues to fascinate and intrigue in equal
measure.

\section*{Acknowledgments}
We thank Kymani Armstrong-Williams, Alfredo Guevara and Ricardo
Monteiro for helpful discussions and / or comments on the
manuscript. We are also grateful to Mariana Carrillo Gonz\'{a}lez and
Justinas Rumbutis for conversations and collaboration on related
topics. This work has been supported by the UK Science and Technology
Facilities Council (STFC) Consolidated Grant ST/P000754/1 ``String
theory, gauge theory and duality'', and by the European Union's
Horizon 2020 research and innovation programme under the Marie
Sk\l{}odowska-Curie grant agreements No. 764850 ``SAGEX'' and
No. 847523 ‘INTERACTIONS’. AL is supported in part by Independent Research Fund Denmark, grant number 0135-00089B. NM is supported by STFC grant
ST/P0000630/1 and the Royal Society of Edinburgh Saltire Early Career Fellowship. We are grateful to the Kavli Institute for
Theoretical Physics for their hospitality at the High-Precision
Gravitational Waves Program, where parts of this work were carried
out.  This research was supported in part by the National Science
Foundation under Grant No. NSF PHY-1748958.  No new data were
generated or analysed during this study.

\bibliography{refs}
\end{document}